\documentclass[12pt]{article}
\usepackage{latexsym,cite,amssymb,amsmath,times}
\usepackage{enumitem}
\usepackage{float}
\input amssym.def
\input amssym.tex
\usepackage{dsfont}

\makeatletter
\renewcommand{\theequation}{\thesection.\arabic{equation}}
\@addtoreset{equation}{section}
\makeatother

\newcommand{\diag}{{\rm diag}}

\newcommand{\half}{{{\textstyle\frac{1}{2}}}}
\newcommand{\quarter}{{{\textstyle\frac{1}{4}}}}

\newcommand{\be}{\begin{equation} }
\newcommand{\ee}{\end{equation} }
\newcommand{\ba}{\begin{array}}
\newcommand{\ea}{\end{array}}

\newcommand{\su}{\mathbf{su}}
\newcommand{\SU}{\mathbf{SU}}
\newcommand{\so}{\mathbf{so}}

\newcommand{\osp}{\mathbf{osp}}

\newcommand{\rON}{\mathbf{O}(\cN)}
\newcommand{\rU}{\mathbf{U}}

\newcommand{\etaodd}{{\cJ}}
\newcommand{\Tw}{T}
\newcommand{\tx}{\tilde{x}}
\newcommand{\spt}{\mathbf{sp}(2,\R)}
\newcommand{\soN}{\mathbf{so}(\cN)}
\newcommand{\OSp}{\mathbf{OSp}}
\newcommand{\OSpN}{{\OSp(\cN|2,\R)}}
\newcommand{\ospN}{{\osp(\cN|2,\R)}}
\newcommand{\Ooo}{{\mathbf{O}(1,1)}}
\newcommand{\ODD}{{\mathbf{O}(D,D)}}

\def\K{{\cal{K}}}

\def\rd{{\rm{d}}}

\newcommand\cC{{\cal C}}
\newcommand\cD{{\cal D}}
\newcommand\cE{{\cal E}}
\newcommand\cF{{\cal F}}

\newcommand\cJ{{\cal J}}

\newcommand\cL{{\cal L}}
\newcommand\cM{{\cal M}}
\newcommand\cN{{\cal N}}

\newcommand\cQ{{\cal Q}}

\newcommand\cV{{\cal V}}

\def\R{{\mathds{R}}}

\def\cQ{{\mathbf{Q}}}

\def\HU1{{H_{{\scriptscriptstyle{\rm u}(1)}}}}

\def\trU1{{\rm tr}_{\scriptscriptstyle{{\rm u}(1)}}}
\def\Tr{{\rm Tr}}

\def\I_M{{I_{\scriptscriptstyle M\times M}}}

\def\6D{{\scriptscriptstyle{6D}}}
\def\4D{{\scriptscriptstyle{4D}}}
\def\3D{{\scriptscriptstyle{3D}}}
\def\2Dstring{{\scriptstyle{2D\,{\rm string}}}}
\def\2DSYM{{\scriptstyle{2D\,{\rm SYM}}}}
\def\SYMQM{{\scriptstyle{{\rm SYMQM}}}}

\def\N{{\cal  N}}
\def\K{{K}}

\def\cL{{{\cal L}}}

\def\dis{\displaystyle}

\begin{document}
\begin{titlepage}
\title{
\vskip 2cm
Superconformal  Yang-Mills quantum mechanics and\\  Calogero model with     $\OSpN$  symmetry}

\author{\sc Neil B. Copland${}^{\sharp}$,   ~Sung Moon Ko${}^{\dagger}$    
~and~  Jeong-Hyuck Park${}^{\dagger}$}
\date{}
\maketitle \vspace{-1.0cm}
\begin{center}
${}^{\sharp}$Center for Quantum Spacetime, Sogang University, Seoul 121-742, Korea\\
${}^{\dagger}$Department of Physics, Sogang University,  Seoul 121-742, Korea\\~\\
\texttt{ncopland@sogang.ac.kr~~sinsmk2003@sogang.ac.kr~~park@sogang.ac.kr}
~{}\\
\end{center}
\begin{abstract}
\vskip0.2cm
\noindent  In  spacetime dimension two,  pure Yang-Mills possesses no physical degrees  of freedom, and consequently  it admits a supersymmetric extension to couple  to an arbitrary number, $\cN$ say, of Majorana-Weyl gauginos. This   results  in    $(\cN,0)$ super Yang-Mills. Further, its dimensional reduction to mechanics  doubles the number of  supersymmetries,  from $\cN$ to ${\cN+\cN}$,  to include conformal supercharges, and leads to a superconformal Yang-Mills quantum mechanics with symmetry group $\OSpN$.  We   comment on its   connection to  $AdS_{2}\times S^{\cN-1}$ and reduction to  a supersymmetric Calogero model. 
 \end{abstract}
 ~\\

{\small
\begin{flushleft}
~~~~~~~~\textit{PACS}:  11.30.Pb,  11.25.Tq 
\end{flushleft}}
\thispagestyle{empty}
\end{titlepage}
\newpage
\tableofcontents 


\section{Introduction and summary}

The $AdS$/CFT correspondence is a holographic duality between string theory in higher dimensional anti-de Sitter space and gauge theory in lower dimensional flat spacetime, the prototypical example being four-dimensional  $\cN{=4}$ supersymmetric  Yang-Mills theory dual to string theory on an $AdS_5\times S^5$ background~\cite{Maldacena:1997re}.  While  other examples in various spacetime dimensions  have been much studied, the lowest dimensional case, \textit{i.e.~}the $AdS_{2}$/CFT$_{1}$ correspondence, is least understood~\cite{Aharony:1999ti}. Although  there has been much work on conformal and superconformal mechanics\cite{hep-th/9911066,Fedoruk:2010bt,Galajinsky:2003cw,Bellucci:2004wn,Galajinsky:2007gm,arXiv:1111.6255,arXiv:1112.0995,1112.1947}, the connection to string theory and supergravity is less clear: we do not have the now familiar picture of the gravity side as the near-horizon geometry of coincident branes with the worldvolume gauge theory living on the boundary of the $AdS$ near-horizon region.\\

Conformal mechanics is not usually formulated as a gauge theory, but as some multi-particle or supersymmetric extension of the original de Alfaro, Fubini and Furlan (DFF) conformal mechanics with inverse $x^2$ potential~\cite{deAlfaro:1976je}  given by
\be
S=\int {\rm{d}}t\,\left(\dot{x}^2-\frac{g}{x^2}\right)\,,
\ee
with $g$ constant. This model is related to the Calogero model of a class of integrable multi-particle systems\cite{Calogero:1969xj, Calogero:1969ie} (see also \cite{Hoppe2004,Hoppe2007}), being obtainable from the two-particle case. Generalizations of these models to higher number of particles, $K$, and supersymmetry were reviewed recently by Fedoruk, Ivanov  and  Lechtenfeld~\cite{1112.1947}  (see also \cite{Hakobyan:2011gj}). Although these models in their final form do not contain gauge fields, there are a class of supersymmetric models that have been  found by gauging models with auxiliary fields in the fundamental representation of a  gauge group\cite{Fedoruk:2008hk,1112.1947} (some early bosonic work used a similar approach \cite{Polychronakos:1991bx}).\\

In much of the discussion of the $AdS_2$/CFT$_1$ correspondence the bulk theory has been considered from a genuine   two-dimensional viewpoint (\textit{e.g.} \cite{Cadoni:2000gm,Astorino:2002bj,Strominger:2003tm, Verlinde:2004gt}). However, it is  also possible   to  embed an $AdS_{2}$  into critical string theory by considering  an extra  factor such as a sphere. Indeed, D0-particles can give rise to $AdS_{2}\times S^8$ geometry (\textit{e.g.} see discussion in \cite{Gibbons:1998fa}). Such geometries with a sphere factor commonly appear in the near-horizon geometry of black holes, as for every known extremal black hole the near-horizon geometry contains an $AdS_2$ factor\footnote{$AdS$ factors can also be seen in black hole moduli spaces\cite{hep-th/9911066,Maloney:1999dv}.}. Compactifying string theory with intersecting D-branes wrapping an internal space gives rise to such black holes, which should be dual to some CFT$_1$ on the uncompactified time direction of the branes' worldvolume intersection. This has been  used from early investigations into the relations with string theory\cite{Strominger:1998yg} to recent investigations into black hole  entropy\cite{Sen:2008vm,Sen:2011cn}.\\

A more explicit realization by Claus {\it et al.}\cite{Claus:1998ts} noted that a worldline action of a superparticle in the $AdS_2\times S^2$ near horizon geometry of an extreme Reissner-Nordstr\"{o}m black hole reduced to superconformal mechanics in a certain limit (see also \cite{deAzcarraga:1998ni,Bellucci:2002va}). This allowed Gibbons and Townsend\cite{Gibbons:1998fa} to argue the CFT$_1$ describing the brane construction of this black hole was an $\N=4$ Calogero model, and further that it should be obtainable from a dimensional reduction of the super Yang-Mills on the two-dimensional intersection of two of the branes (the connection between Calogero models and two-dimensional Yang-Mills having been made before\cite{Gorsky:1993pe}).\\

Pure Yang-Mills theories in spacetime dimensions three, four, six and ten admit  a minimal supersymmetric extension. That is to say,    without introducing additional  bosonic scalar fields,  it is possible to match the bosonic and fermionic  physical degrees of freedom,
\be
\ba{ll}
D=3~~~\,\mathrm{(Majorana~spinor)}:~~&~~3-2=\half\times 2^{1}\,,\\
D=4~~~\,\mathrm{(Majorana~spinor)}:~~&~~4-2=\half\times 2^{2}\,,\\
D=6~~~\,\mathrm{(complex~Weyl~spinor)}:~~&~~6-2=\half\times (\half\times 2^{3}+\half\times 2^{3})\,,\\
D=10~~\mathrm{(Majorana{-Weyl}~spinor)}:~~&~~10-2=\half\times\half\times 2^{5}\,.
\ea
\ee
Technically at the Lagrangian level, the supersymmetry is realized by  virtue of Fierz identities   that cancel the cubic order terms in gauginos arising from the supersymmetric variation of the Yukawa term. Further, the supersymmetry in the above dimensions also has deep connection to the division algebras~\cite{Kugo:1982bn,Evans:1987tm,Evans:2009ed,Baez:2009xt}. \\
\indent On the other hand, in two-dimensional spacetime pure Yang-Mills has no physical degrees of freedom. This hints at the possibility of  a supersymmetric extension coupling to an arbitrary number, $\cN$, of gauginos. The matching of the  physical degrees will then be
\be
2-2=\cN\times 0\,.
\ee
The present work is based on the observation that indeed such a supersymmetric extension is possible. The main contents as well as the organization of this paper are as follows:
\begin{itemize}
\item In section \ref{sec2DSYM} we  explicitly construct $2D$ $(\cN,0)$ super Yang-Mills action, of which the bosonic sector is simply pure Yang-Mills and the fermionic sector consists of an arbitrary number, $\cN$,  of  Majorana-Weyl gauginos. 

\item In section \ref{subsecSCFYMQM}, we perform a dimensional reduction of  $2D$ $(\cN,0)$ super Yang-Mills to a Yang-Mills quantum mechanics. We show that the resulting one-dimensional Yang-Mills is superconformal as  the supersymmetry becomes  doubled, $\cN\rightarrow\cN+\cN$, to include conformal supercharges.

\item In section \ref{subsecDEFORM}, we identify  the   superconformal symmetry group as  $\OSpN$. 

\item Further, in section \ref{subsecSCFS},  we generalize  the Yang-Mills quantum mechanics to include an arbitrary time dependent mass term and a one-dimensional  Chern-Simons term,   without breaking the    superconformal symmetry. In the case when the mass parameter is  constant, the massive model corresponds to the radial quantization of the massless model.

\item In section \ref{secELIMA},  we consider two different gauge choices, one a complete gauge fixing and the other a partial gauge fixing. In the latter case,  we break the gauge group ${\rm{U}}(\K)$ to ${\rm{U}}(1)^{\K}$ by diagonalizing   the unique  bosonic dynamical  matrix  and eliminate the  off-diagonal components of the gauge field using its equation of motion. This results in a Calogero-like model with  inverse square potential. We show that the resultant model maintains ${\cN+\cN}$ superconformal symmetries.   

\item In section \ref{secAdS},  we map the geodesic motion of a point particle on   $AdS_{2}\times S^{\cN-1}$  to the Abelian sector of our  massive super Yang-Mills quantum mechanics, where the mass is given by the inverse of the $AdS_{2}$ radius.

\item Section \ref{secCON} contains  concluding remarks. 
\end{itemize}

The enhancement of  ordinary supersymmetry  to superconformal symmetry  upon  dimensional reduction is well known for $4D$ super Yang-Mills:\footnote{For other examples of the supersymmetry enhancement upon dimensional reduction, see \cite{Hyun:2002fk,Hyun:2003se}.} Dimensional reduction of $10D$ minimal super Yang-Mills to $4D$ $\cN{=4}$  super Yang-Mills  doubles the number of supersymmetries from  sixteen  to thirty two, resulting in the superconformal symmetry, $\SU(2,2|\cN)$.\\\newpage

For earlier  studies on the superconformal mechanics and its super Lie algebra, we refer to  reviews \cite{hep-th/9911066,arXiv:1111.6255,arXiv:1112.0995,1112.1947,Hakobyan:2009ac,Fedoruk:2009pq,Fedoruk:2010wj} and references  therein.  In particular,  in the  $\cN=1,2,4$ cases our model essentially coincides with \cite{1112.1947,Fedoruk:2008hk}.  Compared to them,   the novelties  of our model are  that \textit{i}) it is of Yang-Mills type with gauge group,  $\rU(\K)$,   \textit{ii}) it has an arbitrary number of supersymmetries, $\cN$,  \textit{iii}) it admits an arbitrary time dependent mass deformation as well as a Chern-Simons term,  and \textit{iv}) 
the  corresponding  superconformal group is $\OSpN$. \\

\section{$2D$ $(\cN,0)$ super Yang-Mills\label{sec2DSYM}}
The  Minkowskian $2D$ $(\cN,0)$ super Yang-Mills Lagrangian we propose  is
\be
\cL_{\2DSYM}=\Tr\left[-\quarter F_{\mu\nu}F^{\mu\nu}-i\half\bar{\Psi}_{a}\Gamma^{\mu}D_{\mu}\Psi^{a}\right]\,.
\label{2DSYM}
\ee
Specifically,  we assume  the gauge group, $\rU(\K)$, and  set in a standard manner,
\be
\ba{ll}
F_{\mu\nu}=\partial_{\mu}A_{\nu}-\partial_{\nu}A_{\mu}-i\left[A_{\mu},A_{\nu}\right]\,,~~&~~D_{\mu}\Psi^{a}=\partial_{\mu}\Psi^{a}-i\left[A_{\mu},\Psi^{a}\right]\,.
\ea
\ee
The spinors, $\Psi^{a}$, $a=1,2,\cdots,\cN$, are Majorana-Weyl of definite chirality,
\be
\ba{ll}
\Gamma^{01}\Psi^{a}=+\Psi^{a}\,,~~~&~~~\bar{\Psi}_{a}=\left(\Psi^{a}\right)^{\dagger}\Gamma^{0}=-\bar{\Psi}_{a}\Gamma^{01}\,,
\ea
\label{Weyl}
\ee
such that each  $\Psi^{a}$ has only one Hermitian  spinorial component,
\be
\ba{ll}
\Psi^{a}=\left(\ba{c}2^{-\frac{1}{4}}\psi^{a}\\0\ea\right)\,,~~~&~~~\psi^{a}=\left(\psi^{a}\right)^{\dagger}\,.
\ea
\label{Psipsi}
\ee
Note that both $A_{\mu}$ and $\psi^{a}$ are  $\K\times\K$  Hermitian matrices in the adjoint representation of the gauge group, $\rU(\K)$.

The $(\cN,0)$ supersymmetry transformation is given by 
\be
\ba{ll}
\delta A_{\mu}=i\bar{\cE}_{a}\Gamma_{\mu}\Psi^{a}=-i\bar{\Psi}_{a}\Gamma_{\mu}\cE^{a}\,,~~~&~~~\delta\Psi^{a}=-\half F_{\mu\nu}\Gamma^{\mu\nu}\cE^{a}\,,
\ea
\label{2Dsusy}
\ee
where $\cE^{a}$, $a=1,2,\cdots,\cN$ are   Majorana-Weyl spinorial  Grassmann parameters.

The Lagrangian is invariant up to total derivatives under the supersymmetry transformation. This can be easily seen by employing
the light-cone coordinates,
\be
\ba{llll}
 x^{+}=\textstyle{\frac{1}{\sqrt{2}}}(t+x)\,,~~&~~x^{-}=\textstyle{\frac{1}{\sqrt{2}}}(t-x)\,,~~&~~
  \partial_{+}=\textstyle{\frac{1}{\sqrt{2}}}(\partial_{t}+\partial_{x})\,,~~~~&~~~~
  \partial_{-}=\textstyle{\frac{1}{\sqrt{2}}}(\partial_{t}-\partial_{x})\,.
 \ea
 \ee
Setting the metric to be
\be
ds^{2}=-dt^{2}+dx^{2}=-2dx^{+}dx^{-}\,,
\ee
we choose the gamma matrices as
\be
\ba{ll}
\Gamma^{+}=-\Gamma_{-}=\left(\ba{cc}0&\sqrt{2}\\0&0\ea\right)\,,~~~&~~~
\Gamma^{-}=-\Gamma_{+}=\left(\ba{cc}0&0\\-\sqrt{2}~&0\ea\right)\,.
\ea
\ee
The chiral spinors then satisfy
\be
\ba{lll}
\Gamma_{-}\Psi^{a}=0\,,~~~&~~~\Gamma_{-}\cE^{a}=0\,,~~~&~~~\Gamma^{0}\Gamma^{\mu}D_{\mu}\Psi=-\sqrt{2}\,D_{-}\Psi\,,
\ea
\label{lightchiral}
\ee
which along with (\ref{Psipsi}) reduces the Lagrangian~(\ref{2DSYM}) to
\be
\cL_{\2DSYM}=\Tr\left[\half (F_{+-})^{2}+i\half\psi_{a}D_{-}\psi^{a}\right]\,.
\label{l2d}
\ee
Hereafter $\rON$ indices will be contracted with the Kronecker-delta symbols,  $\delta^{ab}$, $\delta_{ab}$,  using  Einstein convention.  \\
\indent The crucial observation is that only $A_{-}$ couples to the gauginos, while its supersymmetry variation is trivial due to the chirality of spinors (\ref{lightchiral}),
\be
\delta A_{-}=i\bar{\cE}_{a}\Gamma_{-}\Psi^{a}=0\,.
\ee
Thus, unlike higher dimensional super Yang-Mills,  no cubic-order terms in gauginos appear from the  supersymmetry variation of the Yukawa term. Accordingly, there is no call for any Fierz identity for the arbitrary  $(\cN,0)$ supersymmetry to hold!\\

\section{Superconformal Yang-Mills quantum mechanics\label{secSCFYMQM}}
In this section, we first perform a dimensional reduction of the above $2D$ $(\cN,0)$ super Yang-Mills to a $1D$ matrix model  (Yang-Mills quantum mechanics) and   discuss its supersymmetric  deformation. We identify its superconformal  symmetry group as $\OSpN$.

\subsection{Dimensional reduction\label{subsecSCFYMQM}}
The dimensional reduction of the above $2D$ $(\cN,0)$ super Yang-Mills (\ref{l2d}) to the $1D$ light-cone time, $x^{-}\equiv t$,  leads to the following  super Yang-Mills quantum mechanics  (SYMQM)\,,
\be
\cL_{\SYMQM}=\Tr\left[\half(D_{t}X)^{2}+i\half\psi_{a}D_{t}\psi^{a}\right]\,,
\label{SYMQM0}
\ee
where the covariant derivative takes the form,
\be
D_{t}=\partial_{t}-i\left[A\,,~~~\right]\,.
\ee
Note that $X$ is the only bosonic physical variable, the gauge field $A$ is auxiliary,  and the fermions $\psi^{a}$, $a=1,2,\cdots,\cN$ carry no spinorial index.  All the variables,  $X,\,A,\,\psi^{a}$  are $\K\times\K$ Hermitian matrices.  \\
\indent The gauge symmetry is given by, with $g\in\rU(\K)$, 
\be
\ba{lll}
X~\longrightarrow~ gXg^{-1}\,,~&~~\psi^{a}~\longrightarrow~ g\psi^{a} g^{-1}\,,~&~~A~\longrightarrow~ gA g^{-1}-i\partial_{t}gg^{-1}\,.
\ea
\label{gaugesym}
\ee
The supersymmetry transformation inherited  from (\ref{2Dsusy}) reads
\be
\ba{lll}
\delta X=i\psi^{a}\epsilon_{a}\,,~~~&~~~\delta\psi^{a}=D_{t}X\epsilon^{a}\,,~~~&~~~\delta A=0\,.
\ea
\label{ordsusy0}
\ee
Moreover, the super Yang-Mills quantum mechanics enjoys the conformal supersymmetry,
\be
\ba{lll}
\delta^{\prime} X=it\psi^{a}\epsilon^{\prime}_{a}\,,~~~&~~~\delta^{\prime}\psi^{a}=\left(tD_{t}X-X\right)\epsilon^{\prime a}\,,~~~&~~~\delta^{\prime} A=0\,.
\ea
\label{confsusy0}
\ee
~\newline

\subsection{Supersymmetric deformation and superconformal  symmetry\label{subsecDEFORM}}
The above super Yang-Mills quantum mechanics (\ref{SYMQM0})  admits     a supersymmetric mass deformation involving an given arbitrary function of time, without breaking  any of the  ${\cN+\cN}$ supersymmetries,  (\ref{ordsusy0}) and (\ref{confsusy0}), as noted  previously for the case of   $\cN{=1}$~\cite{Park:2005pz}.

After introducing this given arbitrary   function of time, $\Lambda(t)$, which  has  dimension mass squared,   the  deformed superconformal Yang-Mills quantum mechanics is of the general form:
\be
\cL_{\SYMQM}=\Tr\left[\,\half(D_{t}X)^{2}+i\half\psi_{a}D_{t}\psi^{a}+\half\Lambda(t)X^{2}+\kappa A\,\right]\,.
\label{scfYMQM}
\ee
Here we have also added a one-dimensional Chern-Simons term with coefficient (or level) $\kappa$, which must be quantized at  the quantum level~\cite{Bak:2001ze,Nair:2001rt,Park:2001sj,Park:2002eu}.\\

The ``superconformal" symmetry or ${\cN+\cN}$ supersymmetry  transformations are
\be
\ba{lll}
\delta_{+} X=if_{+}\psi_{a}\epsilon^{a}_{+}\,,~~~&~~~
\delta_{+}\psi^{a}=\left(f_{+}D_{t}X-\dot{f}_{+}X\right)\epsilon^{a}_{+}\,,~~~&~~~\delta_{+} A=0\,,\\
\delta_{-} X=if_{-}\psi_{a}\epsilon^{a}_{-}\,,~~~&~~~
\delta_{-}\psi^{a}=\left(f_{-}D_{t}X-\dot{f}_{-}X\right)\epsilon^{a}_{-}\,,~~~&~~~\delta_{-} A=0\,.
\ea
\label{confsusy}
\ee
Here $\epsilon^{a}_{+}$, $\epsilon^{b}_{-}$  are ${\cN+\cN}$  supersymmetry  parameters, and 
$f_{+}(t)$,  $f_{-}(t)$ are the two independent solutions to  the  second-order differential equation:
\be
\ddot{f}_{\pm}(t)=\Lambda(t)f_{\pm}(t)\,.
\ee
From their independence and  $\frac{{\rm d}~}{{\rm d}t}\left(f_{+}\dot{f}_{-}-f_{-}\dot{f}_{+}\right)=0$,   it follows that      $f_{+}\dot{f}_{-}-f_{-}\dot{f}_{+}$ is a non-vanishing  constant.  Without loss of generality we will henceforth normalize it to unity,
\be
\displaystyle{f_{+}(t)\dot{f}_{-}(t)-f_{-}(t)\dot{f}_{+}(t)=1\,.}
\label{normal}
\ee
In the special case of  $\Lambda(t)=0$, we may set $f_{+}=1$, $f_{-}=t$, and (\ref{confsusy}) reduces to (\ref{ordsusy0}), (\ref{confsusy0}).   On the other hand, when $\Lambda$ is non-zero constant we may choose
\be
\ba{lll}
f_{+}(t)=\frac{1}{\sqrt{\Lambda}}\cosh\left(\sqrt{\Lambda}t\right)\,,~~&~~
f_{-}(t)=\frac{1}{\sqrt{\Lambda}}\sinh\left(\sqrt{\Lambda}t\right)~~&~~~\mbox{for~~}\Lambda>0\,,\\
f_{+}(t)=\frac{1}{\sqrt{|{\Lambda}|}}\cos\left(\sqrt{|{\Lambda}|}t\right)\,,~~&~~
f_{-}(t)=\frac{1}{\sqrt{|{\Lambda}|}}\sin\left(\sqrt{|{\Lambda}|}t\right)~~&~~~\mbox{for~~}\Lambda<0\,.
\ea
\ee
~\\

For generic $\Lambda(t)$,  under the superconformal transformations (\ref{confsusy}) the mass deformed  Lagrangian (\ref{scfYMQM}) transforms as a total derivative,
\be
\delta_{\pm}\cL_{\SYMQM}=\frac{\rd~}{\rd t}\Tr\left[D_{t}X\delta_{\pm}X-i\half\psi^{a}\delta_{\pm}\psi_{a}
\right]\,,
\label{totalDeriv}
\ee
ensuring the invariance of the   corresponding action.\\

\indent Further,  the  Lagrangian (\ref{scfYMQM}) possesses  a bosonic $\soN\times\spt$ symmetry, as follows.
\begin{itemize}
\item  $\soN$ rotation, 
\be
\ba{lll}
\delta_{\soN}X=0\,,~~~&~~~\delta_{\soN}\psi^{a}=M^{a}{}_{b}\psi^{b}\,,~~~&~~~
\delta_{\soN}A=0\,,
\ea
\label{soNtr}
\ee
where $M_{ab}=-M_{ba}\in\soN$.
\item $\spt\equiv\so(1,2)\equiv\mathbf{sl}(2,\R)\equiv\su(1,1)$ conformal symmetry,
\be
\ba{lll}
\delta_{\spt}X=\delta tD_{t}X-\half\left(\frac{\rd~}{\rd t}\delta t\right)X\,,~~~&~~~\delta_{\spt}\psi^{a}=0\,,~~~&~~~
\delta_{\spt}A=0\,,
\ea
\label{deltattr}
\ee
where $\delta t$ is a  generic solution to the third order differential equation~\cite{Erdmenger:2006eh},
\be
\displaystyle{\frac{{\rm d}^{3}\delta t}{{\rm d}t^{3}}=4
\Lambda\frac{{\rm d}\delta t}{{\rm d}t}+2\frac{{\rm d}\Lambda}{{\rm d}t}\delta t\,.}
\label{3rdD}
\ee
Note that Eq.(\ref{deltattr})  is consistent with the fact that $X$ has conformal weight   one half. 
In fact,  if we define
\be
\ba{lll}
J_{0}:=i\half\left(f_{+}^{2}+f_{-}^{2}\right)\partial_{t}\,,~~&~~
J_{1}:=i\half\left(f_{+}^{2}-f_{-}^{2}\right)\partial_{t}\,,~~&~~
J_{2}:=if_{+}f_{-}\,\partial_{t}\,,
\ea
\label{Jso(1,2)}
\ee
the three independent solutions of (\ref{3rdD}) can be generated by $J_{\mu}$, $\mu=0,1,2$. \\
\indent Further,  using (\ref{normal}), we obtain the commutator  relations, 
\be
\ba{lll}
{}\left[J_{0},J_{1}\right]=-iJ_{2}\,,~~~&~~~{}\left[J_{1},J_{2}\right]=+iJ_{0}\,,~~~&~~~
{}\left[J_{2},J_{0}\right]=-iJ_{1}\,,
\ea
\ee
which is the Lie algebra $\mathbf{sp}(2,\R)\equiv\so(1,2)\equiv\mathbf{sl}(2,\R)\equiv\su(1,1)$,\textit{\,~i.e.~}the  isometry group of  $AdS_{2}$.  
\end{itemize}
~\\

It is worth noting that when $\Lambda\equiv m^{2}$ is constant,  the mass deformed SYMQM (\ref{scfYMQM}) can be identified as the radial quantization of the massless SYMQM (\ref{SYMQM0}),  for which the time coordinate and the fields need to be  redefined according to
\be
\ba{lll}
\left(\ba{c}
t\\
X(t)\\
A(t)\\
\psi^{a}(t)\ea
\right)~&~\Longrightarrow~&~\left(\ba{c}
\frac{1}{m}\,e^{2mt}\\
\sqrt{2}\,e^{mt}X(t)\\
\half e^{-2mt}A(t)\\
\psi^{a}(t)\ea
\right)\,.
\ea
\ee
Otherwise, \textit{i.e.~}when $\Lambda(t)$ has nontrivial time dependence, the mass deformed SYMQM (\ref{scfYMQM}) cannot be obtained from the field redefinition of the  massless SYMQM.\footnote{This can be traced back to the fact that  the following term is not a total derivative and hence cannot be ignored for generic $\Lambda(t)$, 
\[
\Tr\left(\sqrt{\Lambda(t)}\,X{\frac{\rd~}{\rd t}}X\right)\,.
\]}
~\\

\subsection{Superconformal   group, $\OSpN$\label{subsecSCFS}}
From the ${\cN+\cN}$ supersymmetry  (\ref{confsusy}),    $\soN$  symmetry (\ref{soNtr}) and $\spt$ symmetry  (\ref{deltattr}), along with (\ref{totalDeriv}) and (\ref{Jso(1,2)}), it is straightforward to compute the corresponding Noether charges. We write them in Hamiltonian formalism where the conjugate momentum of $X$ is
\be
P=D_{t}X\,,
\ee
and the Hamiltonian is independent of the fermions, 
\be
H=\Tr\left[\half P^{2}-\half \Lambda(t)X^{2}\right]\,.
\label{Hamiltonian}
\ee
However,  the equation of motion of the auxiliary gauge field $A$ gives rise to a first-class Gauss constraint, 
\be
\left[P,X\right]+i\psi^{a}\psi_{a}-i\kappa=0\,.
\label{Gauss}
\ee
Upon quantization,  with $r,s,t,u$ as  $\K\times\K$ matrix indices, we have 
\be
\ba{ll}
\left[X^{r}{}_{s},P^{t}{}_{u}\right]=i\delta^{r}{}_{u}\delta^{t}{}_{s}\,,~~~~&~~~~
\left\{\psi^{a}{}^{r}{}_{s},\psi^{b}{}^{t}{}_{u}\right\}=\delta^{ab}\delta^{r}{}_{u}\delta^{t}{}_{s}\,,
\ea
\ee
and the  left hand side of the equality in (\ref{Gauss}) corresponds to the $\rU(\K)$ gauge symmetry generator.\footnote{This may involve a  normal ordering prescription  of the expression and a consequent  renormalization of the Chern-Simons level. } We further consider  a change of the bosonic variables, 
\be
\ba{ll}
A_{+}:=f_{+}P-\dot{f}_{+}X\,,~~~&~~~A_{-}:=f_{-}P-\dot{f}_{-}X\,,
\ea
\ee
which satisfies   with  the normalization (\ref{normal}),
\be
\left[A_{+}{}^{r}{}_{s}\,,A_{-}{}^{t}{}_{u}\right]=i\delta^{r}{}_{u}\delta^{t}{}_{s}\,.
\ee
~\\

The Noether charges corresponding to all the symmetries  are then as follows:
\begin{enumerate}
\item Supercharges,
\be
\ba{ll}
Q_{+}^{a}=\Tr\left(\psi^{a}A_{+}\right)\,,~~~&~~~Q_{-}^{a}=\Tr\left(\psi^{a}A_{-}\right)\,.
\ea
\ee
\item $\soN$,
\be
\cM^{ab}=i\,\Tr\left(\psi^{a}\psi^{b}\right)\,.
\ee
\item  $\mathbf{sp}(2,\R)\equiv\so(1,2)\equiv\mathbf{sl}(2,\R)\equiv\su(1,1)$,
\be
\ba{lll}
\cJ_{0}=\half\Tr\left(A_{+}^{2}+A_{-}^{2}\right),~~&~~~
\cJ_{1}=\half\Tr\left(A_{+}^{2}-A_{-}^{2}\right),~~&~~~
\cJ_{2}=\half\Tr\left(A_{+}A_{-}+A_{-}A_{+}\right).
\ea
\ee
\end{enumerate}

In order to write all the super-commutator relations in an $\so(1,2)$ covariant manner, we introduce   three-dimensional  $2\times 2$ gamma matrices, 
\be
\ba{lll}
\gamma^{0}=\left(\ba{rr}0&-1\\1&0\,\ea\right)\,,~~~&~~~
\gamma^{1}=\left(\ba{rr}0\,&-1\\-1&0\,\ea\right)\,,~~~&~~~
\gamma^{2}=\left(\ba{rr}1&0\,\\0&-1\ea\right)\,,
\ea
\ee
satisfying
\be
\ba{ll}
\gamma^{\mu}\gamma^{\nu}+\gamma^{\nu}\gamma^{\mu}=2\eta^{\mu\nu}\,,~~~&~~~
(\gamma^{\mu})^{\alpha}{}_{\beta}(\gamma_{\mu})^{\gamma}{}_{\delta}=2\delta^{\alpha}{}_{\delta}\delta^{\gamma}{}_{\beta}-\delta^{\alpha}{}_{\beta}\delta^{\gamma}{}_{\delta}\,,
\ea
\label{gamma3}
\ee
where $\eta=\diag(-++)$ is the three-dimensional Minkowskian metric.  \\

Further, we combine $Q^{a}_{+}$ and $Q^{a}_{-}$ to form a set of  two-component Majorana spinorial supercharges, $\cQ^{a\alpha}$, $a=1,2,\cdots,\cN$, $~\alpha=1,2$, 
\be
\ba{ll}
\cQ^{a}=\left(\ba{l}Q^{a}_{+}\\ Q^{a}_{-}\ea\right)\,,~~~~&~~~~
\bar{\cQ}_{a}=\cQ^{aT}\gamma^{0}=\left(Q_{-}^{a}\,,\,-Q_{+}^{a}\right)\,.
\ea
\ee
\newpage

It then follows that the symmetry algebra of the  superconformal Yang-Mills quantum mechanics (\ref{scfYMQM}) is the super Lie algebra $\ospN$ which takes the form\footnote{The Jacobi identity of the superalgebra  (\ref{superalgebra}) involving three supercharges holds due to the completeness relation of the gamma matrices in (\ref{gamma3}),
\[
\left[\left\{\cQ^{a\alpha},\cQ^{b\beta}\right\},\cQ^{c\gamma}\right]+\left[\left\{\cQ^{b\beta},\cQ^{c\gamma}\right\},\cQ^{a\alpha}\right]+\left[\left\{\cQ^{c\gamma},\cQ^{a\alpha}\right\},\cQ^{b\beta}\right]=0\,.
\]}

\be
\ba{l}
\left\{\cQ^{a},\bar{\cQ}_{b}\right\}=\delta^{a}{}_{b}\,\gamma^{\mu}\cJ_{\mu}+\cM^{a}{}_{b}\,1\,,\\
\left[\cJ_{\mu},\cQ^{a}\right]=-i\gamma_{\mu}\cQ^{a}\,,\\
\left[\cM^{ab},\cQ^{c}\right]=i\left(\delta^{cb}\cQ^{a}-\delta^{ca}\cQ^{b}\right)\,,\\
\left[\cJ_{\mu},\cJ_{\nu}\right]=-2i\epsilon_{\mu\nu\lambda}\cJ^{\lambda}\,,\\
\left[\cM_{ab},\cM_{cd}\right]=i\left(\delta_{ad}\cM_{bc}-\delta_{bd}\cM_{ac}+\delta_{bc}\cM_{ad}-\delta_{ac}\cM_{bd}\right)\,,
\ea
\label{superalgebra}
\ee
where   $\epsilon_{\mu\nu\lambda}$ denotes  the usual  totally anti-symmetric tensor with $\epsilon_{012}\equiv 1$.\\

Finally,  the Casimir  of the $\ospN$ superalgebra is
\be
\ba{ll}
\cC_{\ospN}=\cJ_{\mu}\cJ^{\mu}+\half\cM_{ab}\cM^{ab}-i\bar{\cQ}_{a}\cQ^{a}\,,~~~&~~~
\dis{\Big[\,\cC_{\ospN}\,,\,\mbox{anything}\,\Big]=0\,.}
\ea
\label{Casimirosp}
\ee
~\\
\newpage

\section{Reduction to  supersymmetric  Calogero models\label{secELIMA}}
In this section,  we discuss  two possible  gauge fixings which  allow  us to write down two  types of     supersymmetric generalization of the Calogero model  having  ${\N+\cN}$ superconformal symmetries.

\subsection{Type I: Complete  gauge fixing, $X$ diagonal and $A$ off-diagonal\label{subseccomplete}}
We recall   the action (\ref{scfYMQM}),
\be
\cL_{\SYMQM}=\Tr\left[\half(D_{t}X)^{2}+i\half\psi_{a}D_{t}\psi^{a}+\half\Lambda(t)X^{2}+\kappa A\right]\,.
\label{scfYMQMCS}
\ee
We use the $\rU(K)$ gauge symmetry to fix $X$ to be diagonal,
\be
\ba{lll}
X_{ij}=0~~&~~\mbox{for}~&~i\neq j\,.
\ea
\label{diagonalX}
\ee
This still leaves a $\rU(1)^{\K}$ diagonal residual gauge symmetry which can be used to eliminate the diagonal components of $A$,
\be
\ba{ll}
A_{ii}=0~~&~~\mbox{(no~sum)}\,.
\ea
\label{offdiagonalA}
\ee
The supersymmetry transformations (\ref{confsusy}) do not preserve these gauge fixings, (\ref{diagonalX}) and (\ref{offdiagonalA}),  and must be compensated by an additional gauge transformation to bring us back to the gauge choice. We let 
\be
\ba{ll}
x_i=X_{ii}~~&~~\mbox{(no~sum)}\,,
\ea
\ee
and, since under a gauge transformation $\delta X=i[\lambda, X]$, we require
\be
\delta_{\pm} X_{ij}=if_{\pm}\psi_{a\, ij}\epsilon^{a}_{\pm}+i\lambda_{ij}(x_j-x_i)=0\qquad\mbox{for}\qquad i\neq j\, ,
\ee
and in order to also preserve the off-diagonal form of $A$ (\ref{offdiagonalA})  we must choose the compensating gauge symmetry parameter as
\be\label{preservegauge}
\lambda_{ij}=\left\{
\begin{array}{l}
\frac{1}{x_i-x_j}f_\pm \psi_{a\, ij}\varepsilon^a_{\pm}~\,\qquad\mbox{for}\qquad i\neq j\,,\\
i\dis{\int \rd t~ [A,\lambda]_{ii}} \quad\qquad\mbox{for}\qquad i=j\,.
\end{array}
\right.
\ee
Note that the diagonal part of $\lambda$ drops out of the commutator, $ [A,\lambda]_{ii}$ in the second line. We then have the Lagrangian,
\be\label{diagaction}
\cL=\half\dot{x}_i^2+\half(x_i-x_j)^2A_{ij}A_{ji}+i\half\psi_{aij}\dot{\psi}_{ji}^a+\psi_{a\,ij}A_{jk}\psi^a_{ki}+\half\Lambda(t) x_i^2\, ,
\ee
where $A$ is strictly  off-diagonal (\ref{offdiagonalA}) and repeated indices are summed over. Defining the combination of supersymmetry and gauge transformation (\ref{preservegauge})
\be
\delta'_{\pm}=\delta_{\pm}+\delta_{\lambda}\, ,
\ee
then gives $\N+\N$ supersymmetries of the Lagrangian (\ref{diagaction}).  The original $\rU(K)$ gauge symmetry is now  completely broken.\\~\\


\subsection{Type II: Super Calogero model  with $\rU(1)^{K}$ unbroken  gauge symmetry}
In this subsection, we also take the diagonal gauge for $X$ (\ref{diagonalX}) but do not  impose (\ref{offdiagonalA}).  
Since the gauge field, $A$, is auxiliary we can eliminate it from the Lagrangian using its equation of motion, which is, from the Gauss constraint (\ref{Gauss}), given by
\be
(x_i-x_j)^2A_{ij}=\half\{\psi^a,\psi_a\}_{ij}-\kappa\delta_{ij}\,.
\ee
However, the diagonal components, $a_i=A_{ii}$ (no sum), are not specified by this constraint, and we choose not to gauge them away as was done in the previous  subsection. For the time being  they remain as auxiliary gauge fields for the unbroken gauge group,  $\rU(1)^{K}$.\\~\\

\newpage

After eliminating the off-diagonal components of $A$,  we have a Calogero model type  Lagrangian,
\be\label{diagactionnoA}
\textstyle{\cL=\half\dot{x}_i^2+i\half\psi_{aij}\dot{\psi}_{ji}^a+\half\Lambda(t)x_i^2-\sum_{i\neq j}{\frac{\{\psi_{a},\psi^a\}_{ij}\{\psi_{b},\psi^b\}_{ji}}{8(x_i-x_j)^2}}+\sum_ia_i\left(\kappa-\half\left\{\psi_{a},\psi^a\right\}_{ii}\right)\, ,}
\ee
which enjoys   ${\cN+\cN}$   superconformal symmetries,
\be\label{diagxasusy}
\ba{rll}
\delta x_i&={}&if_\pm\psi_{a\, ii}\epsilon_\pm^a\, ,\\
\delta\psi^{a}_{ij}&={}&\left\{\begin{array}{ll}
(f_\pm\dot{x}_i-\dot{f}_\pm x_i)\epsilon^a_\pm& \mbox{for~~}i=j\, ,\\
if_\pm\left[\sum_{k}\frac{
\psi_{b\,ik}\psi^b_{kj}\epsilon^a_\pm}{x_i-x_j}
+\sum_{k\neq i}\frac{\psi^a_{kj}\psi_{b\,ik}\epsilon^b_\pm}{x_i-x_k}
-\sum_{k\neq j}\frac{\psi^a_{ik}\psi_{b\,kj}\epsilon^b_\pm}{x_k-x_j}\right]&\mbox{for~~}i\neq j\, ,
\end{array}\right.\\
\delta a_{i}&={}&\sum_{j\neq i}\sum_{k}\frac{if_\pm (\psi_{a\,ij}\psi_{b jk}\psi^b_{ki}+ \psi_{a\,ji}\psi_{b\,ik}\psi^b_{kj})\epsilon^a_\pm}{(x_i-x_j)^3}\, .
\ea
\ee

We see that $a_i$  is a  Lagrange multiplier for the constraint (with $K$ components),
\be\label{kappaconstraint}
\half\{\psi_a,\psi^a\}_{ii}=\kappa\, ,\qquad \mbox{no sum on~~}i.
\ee
If we integrate it out, then the constraint is needed to show the invariance of the action  under the  supersymmetries   (\ref{diagxasusy}). Similarly, if we had used the $U(1)^K$ residual gauge symmetry to set the $a_i$ to zero ---as done in the previous subsection--- and eliminated  the off-diagonal components of  $A$ from  the action (\ref{diagaction})  by its equation of motion, we would still need the constraint (\ref{kappaconstraint}) in order  to show that the  action is supersymmetric. Of course imposing the constraint {via} a Lagrange multiplier just returns us to the action (\ref{diagactionnoA}).\\

Unlike many known supersymmetric extensions of the multi-particle Calogero model, \textit{e.g.} the original  $\N=2$ Freedman-Mende model~\cite{Freedman:1990gd}, this has $K^2$ fermionic components compared with $K$ bosonic (as was also the case with $\cN=1,2,4$~\cite{1112.1947,Fedoruk:2008hk}). Also, our model exists for arbitrary  number of fermions and supersymmetries,  and for arbitrary  $\Lambda(t)$.  \\

\newpage

\section{Connection to $AdS_{2}\times S^{\cN-1}$\label{secAdS}}
In this section, we discuss the connection of  the Abelian sector  of our superconformal Yang-Mills quantum mechanics to the geodesic motion on $AdS_{2}\times S^{\cN-1}$ spacetime. \\

We begin by considering the  standard global metric of  $AdS_{2}$ with  radius $R$,
\be
ds_{AdS_{2}}^2=R^2(-\cosh^2\rho\, d\tau^{\prime\, 2}+d\rho^2)\, , \qquad \rho\geq0\, .
\ee
We perform a coordinate transformation   following  \cite{Erdmenger:2006eh}, from  $(\rho,\tau^{\prime})$ to $(X,\tau)$ by 
\be
\cosh^2\rho=\frac{1}{1-(X/R)^2}\, , \quad\quad \tau=R\tau'\, ,
\ee
to obtain a new metric,
\be
ds^2_{AdS_2}=-\frac{d\tau^2}{1-(X/R)^2}+\frac{dX^2}{[1-(X/R)^2]^2}\, , \qquad 0\leq X<R\, ,
\ee
in terms of the dimensionful variables $X$ and $\tau$.    We use this for the metric of  $AdS_{2}\times S^{\cN-1}$,
\be
ds^2_{AdS_2\times S^{\cN-1}}
=-\frac{d\tau^2}{1-(X/R)^2}+\frac{dX^2}{[1-(X/R)^2]^2}+g_{\alpha\beta}(\theta)d\theta^{\alpha} d\theta^{\beta}\,,
\ee
where  $\theta^{\alpha}$, $\alpha=1,2,\cdots,{\cN-1}$ are the angular coordinates of $S^{\cN-1}$. \\

After taking the temporal gauge to identify $\tau$  as the worldline coordinate, the point-particle or D$0$ action on $AdS_2\times S^{\cN-1}$ background reads 
\be
\cL=-m\sqrt{[1-(X/R)^2]^{-1}-[1-(X/R)^2]^{-2}\dot{X}^{2}-g_{\alpha\beta}(\theta)\dot{\theta}^{\alpha}\dot{\theta}^{\beta}}\,,
\label{Lag}
\ee
where $m$ is the mass of the particle.\\

The corresponding Hamiltonian is
\be
H=\sqrt{[1-(X/R)^2]P_{X}^{2}+[1-(X/R)^2]^{-1}\left[m^{2}+g^{\alpha\beta}(\theta)P_{\alpha}P_{\beta}\right]}\,,
\label{Ham}
\ee
where $P_{X}$ and $P_{\alpha}$ are the canonical momenta  for $X$ and $\theta^{\alpha}$,
\be
\ba{ll}
P_{X}=H[1-(X/R)^2]^{-1}\dot{X}\,,~~~&~~~P_{\alpha}=H[1-(X/R)^2]g_{\alpha\beta}(\theta)\dot{\theta}^{\beta}\,.
\ea
\ee
The Hamiltonian and the Lagrangian   satisfy
\be
H=-\frac{m^{2}}{\,\cL[1-(X/R)^2]\,}\,.
\ee
Clearly from the form of (\ref{Ham}),  $g^{\alpha\beta}(\theta)P_{\alpha}P_{\beta}$ (the squared   angular momentum)  and the  Hamiltonian itself (the  energy) are conserved quantities, since they `commute' with the Hamiltonian.  From this, one can show straightforwardly   that  the Hamiltonian equation, $\dot{P}_{X}=-\frac{\partial~}{\partial X}H$,  leads to a simple  harmonic oscillatory  motion\footnote{For a recent discussion on  the connection between  harmonic oscillators  and $AdS$ space, see \cite{Freivogel:2011xc}.} with frequency $R^{-1}$, 
\be
\ddot{X}+R^{-2}X=0\,.
\ee
Furthermore, since the  $\so(\cN)$ isometry of the sphere $S^{\cN-1}$ gives rise to the Noether symmetry of the action (\ref{Lag}), there are $\half\cN(\cN-1)$ conserved angular momenta. \\

Thus,  the Abelian sector of the  massive super Yang-Mills quantum mechanics (\ref{scfYMQM}) with the choice of $\Lambda=-R^{-2}$ describes the geodesic motion of a point-particle on $AdS_{2}\times S^{\cN-1}$, where the conserved $\half\cN(\cN-1)$ angular momenta are mapped to the bi-fermionic conserved quantities, $\Tr(\psi_{[a})\Tr(\psi_{b]})$.\\

\newpage

\section{Comments\label{secCON}}
As  generically non-Abelian Yang-Mills quantum mechanics   describes many D0-branes including their interactions, and specifically  the  Abelian sector of our massive super Yang-Mills quantum mechanics can be mapped to the geodesic motion of a single  point-particle on $AdS_{2}\times S^{\cN-1}$,  it seems natural to expect  that the   massive super Yang-Mills quantum mechanics (\ref{scfYMQM}) with a constant mass parameter  $\Lambda=-R^{-2}$  and gauge group $\rU(K)$ provides a worldline  description of  $K$  D0-branes on $AdS_{2}\times S^{\cN-1}$.  Also, in view of the $AdS_{2}$/CFT$_{1}$ correspondence, it is desirable to investigate the non-Abelian nature of the  massive super Yang-Mills quantum mechanics in order to see any `stringy' features.  A recent formal prescription for  the computation of the correlation functions in conformal mechanics~\cite{Chamon:2011xk, Jackiw:2012ur} may help in this direction. \\

Also in the context of  the $AdS_{2}$/CFT$_{1}$ correspondence, it would  be interesting to compute the partition function of  the massive  super Yang-Mills quantum mechanics, $\Tr(e^{-\beta H})$, with $\Lambda=-R^{-2}$.  For the bosonic case of ${\cN=0}$, the quantum eigenstates  are basically generated  by acting with products of  $\Tr(\bar{C}^{n})$ on the quantum vacuum, where  $\bar{C}=\frac{1}{\sqrt{2}}(P\sqrt{R}+iX/\sqrt{R})$ is the matrix valued creation operator and $n=1,2,\cdots, K$. This gives the partition function (with $q=e^{-\beta/R}$)
\be
\dis{\Tr(e^{-\beta H})=q^{\frac{1}{2}K^{2}}\prod_{n=1}^{K}\frac{1}{1-q^{n}}\,,}
\label{Z}
\ee
which agrees with the partition function of  $K$ bosonic harmonic oscillators~\cite{Auluck,Park:2008sk}. Further, in the large $K$ limit or the planar limit,  up to the renormalization of the overall factor, it converges to  the inverse of the  Dedekind eta function (which is  a common special function in the computation of string theory partition functions).  For the path integral derivation of  the formula (\ref{Z}) see  \cite{Kazakov:2000pm} and for the case of $\cN=2$ we refer to \cite{Trzetrzelewski:2007rr,Michishita:2011zz}.  For generic $\cN$ it is an open problem.  \\

Another  stringy feature of the  superconformal Yang-Mills quantum mechanics, at least the massless case (\ref{SYMQM0}),  is that, it can be reformulated as `double field Yang-Mills theory'~\cite{Jeon:2011kp}  to manifest $\mbox{O}(1,1)$ T-duality. We show this in the Appendix.  One more   stringy or $\cM$-theoretic interpretation of our model is that  it may correspond to a  matrix regularization of a membrane worldvolume action, where the two spatial worldvolume directions are replaced by  matrix 
indices~\cite{deWit1988ig,Banks:1996vh} (see also \cite{Kim:2006wg,Park:2008qe} for further discussion). \\

In general, the universal enveloping algebras of   $\so(2,D-1)$  and of the Heisenberg algebra ($[a,\bar{a}]=1$) correspond  to  the  $D$-dimensional  higher spin algebra and the $W_{\infty}$ algebra  respectively.  Since quadratic powers of  Heisenberg algebra generators ($a^{2}$, $\bar{a}^{2}$, ${a\bar{a}+\bar{a}a}$) form the  Lie algebra $\so(2,1)$,  the two-dimensional  higher spin algebra is a subalgebra of the $W_{\infty}$ algebra. In fact, with a $Z_{2}$-grading the $W_{\infty}$  algebra can be identified as the universal enveloping algebra of the super Lie algebra $\osp(1|2)$, where the $Z_{2}$-grading distinguishes the even and odd powers in $a$, $\bar{a}$~\cite{Pope:1989ew,INS-772,INS-783,Bergshoeff:1990cz,Bergshoeff:1991dz,Vasiliev:1995dn,arXiv:0704.0898}.  Our superconformal Yang-Mills  quantum mechanics then generalizes the  $W_{\infty}$ algebra\footnote{The connection between $W_{\infty}$ and abelian conformal  mechanics (DFF) was previously discussed in \cite{Avan:1991kq, Cadoni:2000iz} as well as some superextensions in \cite{Cacciatori:1999rp}.} as well as the the two-dimensional  higher spin algebra in two ways: firstly  it is supersymmetric, containing an arbitrary number, $\cN$, of fermions, with the super Lie  symmetry algebra,  $\osp(\cN|2)$. Secondly  it is non-Abelian such that there are ordering issues; \textit{e.g.~}in general,  $\Tr(A^{2})\neq\Tr(A)^{2}$, $\Tr(A\Psi A\Psi)\neq \Tr(A^{2}\Psi^{2})$ \textit{etc.} More precise identification of the generalized algebra is desirable. \\
~\\
~\\


\begin{center}
\large{\textbf{Acknowledgments}}
\end{center}
We wish to thank Xavier Bekaert, Yoji Michishita,   Armen Nersessian, Dimitri Polyakov and  Richard Szabo for helpful comments. The work was supported by the National Research Foundation of Korea\,(NRF) grants  funded by the Korea government\,(MEST) with the Grant No.  2005-0049409 (CQUeST)  and No.  2010-0002980. \\

\newpage
\appendix
\begin{center}
\large{\textbf{Appendix}}
\end{center}
\setcounter{equation}{0}
\renewcommand{\theequation}{A.\arabic{equation}}
\section{$\Ooo$ T-duality covariant double field  formulation}
String theory possesses   T-duality and     imposes  $\ODD$ structure  on its $D$-dimensional  low energy effective actions~\cite{Buscher:1985kb,Buscher:1987sk,Buscher:1987qj,Giveon:1988tt}.  The $\ODD$ T-duality   can be manifestly  realized   if we  formally double  the spacetime dimension,   from  $D$ to $2D$,  with coordinates,  $x^{\mu}\rightarrow y^{A}=(\tx_{\mu},x^{\nu})$, and reformulate the $D$-dimensional    effective action in terms of $2D$-dimensional language \textit{i.e.}~tensors equipped with  $\ODD$  metric, 
\be
\etaodd_{AB}={{{{\left(\ba{cc}0&1\\1&0\ea\right)}}\,.}}
\label{ODDeta}
\ee
This kind of reformulation  was coined   Double Field Theory (DFT)~\cite{Hull:2009mi,Hull:2009zb,Hohm:2010jy,Hohm:2010pp}.

The new coordinates, $\tx_{\mu}$,  may  be viewed  as the canonical conjugates of the winding modes of closed strings~\cite{Tseytlin:1990nb,Tseytlin:1990va,Siegel:1993xq,Siegel:1993th}.    However,   in DFT,    as a  field theory counterpart to  the level matching condition of closed string theories,    it is required that all the  fields   as well as  all of their possible products should be  annihilated by the $\ODD$ d'Alembert operator, $\partial^{2}=\partial_{A}\partial^{A}$,
\be
\ba{ll}
\partial^{2}\Phi\equiv 0\,,~~~~&~~~~
\partial_{A}\Phi_{1}\partial^{A}\Phi_{2}\equiv 0\,.
\ea
\label{constraint}
\ee
Hence locally, up to $\ODD$ rotation, all the  fields are independent of the dual coordinates~\cite{Hohm:2010jy},
\be
\frac{\partial~~}{\partial\tx_{\mu}}\equiv0\,,
\label{constraint2}
\ee
and   the theory is not truly doubled. \\

With the spacetime dimension formally doubled  in double field theory,   T-duality is   realized   by an $\ODD$ rotation which  acts on the $2D$-dimensional vector  indices of an $\ODD$ covariant tensor  in a standard manner, 
\be
\ba{ll}
\Tw_{A_{1}A_{2}\cdots A_{n}}~~\longrightarrow~~
M_{A_{1}}{}^{B_{1}}M_{A_{2}}{}^{B_{2}}\cdots M_{A_{n}}{}^{B_{n}}\Tw_{B_{1}B_{2}\cdots B_{n}}\,,
~~~&~~~M\in\ODD\,,
\ea
\ee
where the $\ODD$  group is defined by the invariance of the   metric (\ref{ODDeta}),
\be
M_{A}{}^{C}M_{B}{}^{D}\etaodd_{CD}=\etaodd_{AB}\,.
\ee

While the original double field theory focused on the closed string  bosonic effective actions~\cite{Hull:2009mi,Hull:2009zb,Hohm:2010jy,Hohm:2010pp}, the understanding of the underlying differential geometry~\cite{Jeon:2010rw,Jeon:2011cn}  made it possible to construct double field Yang-Mills theory~\cite{Jeon:2011kp} as well as  to include fermions~\cite{Jeon:2011vx,Jeon:2011sq}.\footnote{For related yet inequivalent works, see \textit{e.g.~}\cite{Hohm:2011si,Hohm:2010xe,Hohm:2011ex,Hohm:2011nu}.}\\

 In the current ${D=1}$ case,  the double field Yang-Mills theory and the $\ODD$ covariant Dirac operators get greatly simplified due to the absence of the Kalb-Ramond field, spin connections and the Yang-Mills field strength.  Further, the most general form of an  $\mathbf{O}(1,1)$ group element is given by  the following simple $2\times 2$ matrix,
\be
M_{A}{}^{B}=\pm\left(\ba{cc}0&e^{\phi}\\
e^{-\phi}&0\ea\right)\,.
\label{O11}
\ee
In terms of an einbein, $e$,  the `DFT-vielbein'~\cite{Jeon:2011cn} is given by 
\be
V^{A}=\textstyle{\frac{1}{\sqrt{2}}}
\left(\ba{c}
e\\
e^{-1}
\ea
\right)\,.
\ee
This generates a pair of projections,
\be
\ba{ll}
P^{AB}=V^{A}V^{B}\,,~~~~&~~~\bar{P}^{AB}=\cJ^{AB}-P^{AB}\,,
\ea
\ee
satisfying
\be
\ba{llll}
P^{A}{}_{B}P^{B}{}_{C}=P^{A}{}_{C}\,,~~~&~~~\bar{P}^{A}{}_{B}\bar{P}^{B}{}_{C}=\bar{P}^{A}{}_{C}\,,~~~&~~~P^{A}{}_{B}\bar{P}^{B}{}_{C}=0\,,~~~&~~~P^{A}{}_{B}+\bar{P}^{A}{}_{B}=\delta^{A}{}_{B}\,.
\ea
\ee
The DFT-vielbein is in the fundamental representation of the  $\mathbf{O}(1,1)$  T-duality group, and hence from (\ref{O11}), the $\mathbf{O}(1,1)$  T-duality simply scales the einbein.\\

Now, following \cite{Jeon:2011kp},  we introduce the two-component $\Ooo$ Yang-Mills vector potential,
\be
\cV_{A}=\left(\ba{c}
X\\
A-e^{2}X
\ea
\right)\,,
\ee
where $X$ and $A$ are the dynamical and auxiliary fields in the super Yang-Mills quantum mechanics (\ref{SYMQM0}).\\

After the above $\Ooo$ covariant  reorganization of all the field  variables, using (3.17) of \cite{Jeon:2011kp} and (4.51) of \cite{Jeon:2011vx},   the super Yang-Mills quantum mechanics (\ref{SYMQM0}) with added auxiliary einbein can be reformulated  as a ${D=1}$ supersymmetric double field Yang-Mills theory,
\be
\cL_{\SYMQM}=\Tr\left[P^{AB}\bar{P}^{CD}\cF_{AC}\cF_{BD}
+i\half\psi_{a}V^{A}\cD_{A}\psi^{a}\right]\,,
\ee
which now manifests the $\Ooo$ T-duality structure.\\

%

\end{document}